\documentclass[mathleft, 
]{an}
\usepackage{graphicx}
\usepackage{amssymb,amsmath}
\usepackage{times}
\begin{document}

\Pagespan{789}{}
\Yearpublication{2006}%
\Yearsubmission{2005}%
\Month{11}%
\Volume{999}%
\Issue{88}%

\title{Magneto-rotational and thermal evolution of young neutron stars}
\author{S.B. Popov\inst{1}\fnmsep\thanks{Corresponding author:
  \email{sergepolar@gmail.com}\newline}
}
\titlerunning{Evolution of near-by young neutron stars}
\authorrunning{S. B. Popov}
\institute{
Lomonosov Moscow State University, Sternberg Astronomical Institute,
Universitetski pr. 13,
119991 Moscow, Russia}

\received{30 May 2005}
\accepted{11 Nov 2005}
\publonline{later}


\abstract{%
After a brief review of population synthesis of close-by cooling neutron
stars, I focus on the interpretation of dichotomy of spin
periods of near-by coolers. The existence of
two well separated groups -- short period ($\sim$0.1-0.3 s)
radio pulsars and long period ($\sim$3-10 s) radio quiet sources, aka the
Magnificent seven, -- can not be easily explained in unified models
developed recently (Popov et al. 2010, Gull{\'o}n et al. 2014). I speculate
that the most natural solution of the problem can be in bimodal initial
magnetic field distribution related to the existence of an additional
mechanism of field generation in magnetars.}
\maketitle

\section{Introduction}

 Study of neutron stars (NSs) opens unique opportunities to probe regimes
where matter is placed under extreme conditions: 
very high denstity, strong gravity,
huge electro-magnetic fields, etc.  (see reviews and references in
\cite{pot2014, web2014, man2015}).

 It is possible to confront 
observational data vs. theoretical explanations and 
predictions in several ways. Roughly, they can be divided in two
categories. In the first,
comparison is made using individual sources, in another
-- population analysis is used, 
when general properties of a large set of objects are considered together. 
Typically, the former can give a possibility to go into
tiny details. However, some features of the general picture can be
missed, and so the population approach is also necessary. 
The two methods perfectly complement each others.

In this paper I
discuss population synthesis of young isolated NSs. Some applications of
this method allow us to probe interesting physics of compact objects.
In Sect.~2 population synthesis of isolated NSs is very briefly summarized,
emphasizing our results on cooling compact objects.
Then I discuss joint analysis of several populations of young NSs: radio
pulsars, cooling NSs, and magnetars. 
Sect.~4 provides a description of some new results 
related to an unexplained feature of the population
of near-by cooling isolated NSs. In the last two sections  
a brief duscussion and conclusions are given.


\section{Population synthesis of young NSs}

The population synthesis is a powerful technique, and it is widely used in
astrophysics, including NS studies (\cite{pp2007}).  Historically, first
models were devoted to analysis of radio pulsars (see recent studies for
example in \cite{fgk2006, ip2014, gullon} and references therein to earlier
papers). Starting from late 90s we
developed several population models for different types of isolated 
compact objects. In this section I briefly summarize our results on near-by
cooling NSs. 

 The population of near-by cooling young isolated NSs was mainly discovered with
ROSAT --- the German satellite which produced an all-sky survey in soft
X-rays (see a review in \cite{t2009}). 
There are just about a dozen of such sources. They are either radio
pulsars (including Geminga and a geminga-like source), or radio silent
(see \cite{kond} about deep limits on the radio emission of these sources) 
NSs known as the Magnificent Seven (M7). In the
end of 90s the number of these sources was a big puzzle: 
it was not consistent with general radio pulsar statistics in the Galaxy 
(\cite{nt1999}). Solution was found by Popov et al. (2003)\nocite{p2003}. 

 We demonstrated that the main contribution to the local population of young
NSs is done by the Gould Belt. This is a disc-like 
structure with the size $\sim0.5$~--~0.7~kpc inclined to the Galactic plane. 
It consists of OB-associations with ages from few up to 30-50 Myrs. 
Occasionally, the Sun is situated inside this structure. About 2/3 of young
compact objects inside $\sim 0.6$~kpc are genetically related to the Belt. 

 Later detailed studies of OB-associations distribution in the Belt 
and behind allowed us
to explain the sky distribution of young cooling NSs and make preductions
for future searches (\cite{p2008}). 

 If we neglect (see below about an alternative approach) an 
additional heating
of compact objects due to magnetic field decay, then all astrophysical 
parameters of the scenario are relatively well known (\cite{p2005}). This
allows us to probe the main physical uncertainty -- the rate of NS cooling
related to properties of NS interiors. 
To do it we developed a test of theoretical cooling curves based on
population synthesis of near-by NSs, and applied it to two large sets of
thermal histories for hadron (\cite{p2006a}) and hybrid (\cite{p2006b})
compact stars.

 It was shown that the population synthesis of near-by compact objects can be a
powerful additional test for theoretical models of NS cooling. 
However, in these studies we neglected 
the additional heating due to the magnetic field
decay. For standard radio pulsar fields ($\sim 10^{12}$~G) this assumption
is valid, however for magnetars or their descendants -- it is not so.
Calculations with account for this effect are presented in the next
section.  

\section{Advanced population synthesis of NSs}

Usually in population synthesis studies just one subpopulation of a wide
class of
astronomical objects is modeled. For example, in case of NSs it can be a
separate study of normal radio pulsars, or a separate analysis of millisecond
pulsars, or (as discussed in the previous section) synthesis of only near-by
cooling NSs. In (\cite{p2010}) we made an attempt to model several
subpopulations of compact objects in one framework based on the scenario of
decaying magnetic field developed by the spanish group leaded by Jose Pons
(\cite{aguilera}).   

We used three different codes to model evolution of cooling NSs in the solar
proximity, normal radio pulsars, and magnetars. Different distributions have
been calculated to compare model predictions with observations. In this
study all
three approaches have been unified by the same set of initial conditions and
basic evolutionary laws. In particular, we used a unique single mode Gaussian
distribution for the initial magnetic field of all NSs under study.

 Joint analysis of three sets of results allowed us to derive initial
distributions for magnetic field and spin period. The best fit was obtained
for the initial field distribution with
$\langle \log B_0/\mathrm{[G]} \rangle \approx 13.25$,
$\sigma_\mathrm{B_0}\approx 0.6$, and for the initial period distribution
with 
$\langle p_0  \rangle \approx 0.25$~s, $\sigma_\mathrm{p_0}\approx 0.1$. 

For the first time a population synthesis model of
three different subpopulations of NSs generated from a unique single-mode
initial distribution was shown to be in good correspondence with a large set
of observational data.
Still, 
despite significant progress in realizing the research program known as
``Grand unification for NSs'' (\cite{kaspi2010}), 
there is one feature of the observed
population of near-by cooling NSs which we failed to reproduce in our
calculations.

\section{``One second'' problem}

In this section we describe an unexplained dip in the spin period distribution
of near-by coolers, and discuss some possible solutions of this problem.

\subsection{The origin of the problem}

\begin{figure}
\includegraphics[width=\hsize]{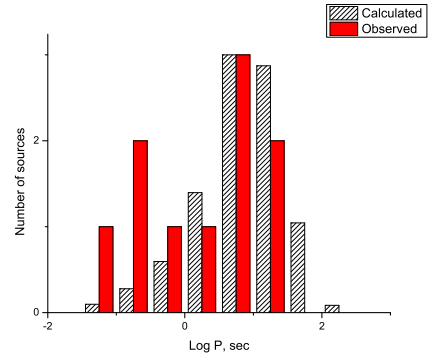}
\caption{Observed and calculated period distributions for close-by young
cooling NSs with ROSAT count rates $>$0.1 cts~s$^{-1}$. 
Calculations are done with the model presented in Popov et al.
(2010)\nocite{p2010}. 
The observed distribution has a bimodal structure, 
while the calculated one has just one peak at $p\sim 10$~s.
Numbers of sources are equal in two distributions.
}
\label{per2010}
\end{figure}

 Close-by cooling NSs are divided into two main groups: radio pulsars and the
M7. They have different spin periods: $\sim 0.1$~--~$0.3$~s 
for pulsars and $\sim 3$~--~$10$~s
for the M7 (see Fig.\ref{per2010}). No sources are
observed between $\sim 0.3$ and $\sim$3~s.  
Oppositely, the calculated distribution is rather smooth.
I.e., no sources with spin periods
$\sim 1$~s $\pm$ half order of magnitude are observed, but the model
predicts them. So, I call it the ``one second problem''. 

In Fig.\ref{per2010} I present results of calculations
for the best model from (\cite{p2010}). 
In this calculations we were not able
to specify the initial period distribution, 
it has been just assumed that periods are
very short. 
However, for our discussion such simplification is acceptable.

\begin{figure}
\includegraphics[width=\hsize]{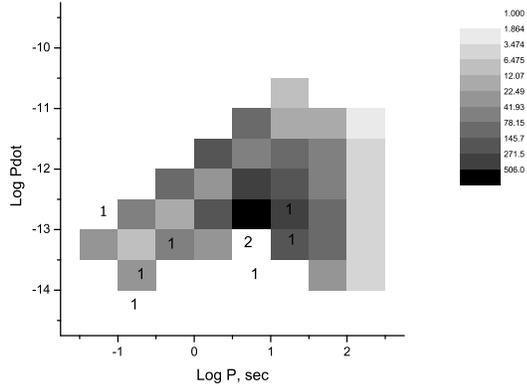}
\caption{ The $p$~--~$\dot p$ distribution for close-by young
cooling NSs with ROSAT count rates $>0.1$~cts~s$^{-1}$. Grey scale
represents
results of our calculations with the evolutionary model described in Popov
et al. (2010)\nocite{p2010}. 
Total amount of calculated sources was not normalized.   
Numbers correspond to the data on real sources (four radio pulsar-like
sources, and five of the M7, for which $p$ and $\dot p$ are well-measured.
}
\label{ppdot2010}
\end{figure}

The problem can be better illustrated with the $p$~--~$\dot p$ plot.
In Fig.~\ref{ppdot2010} I present this diagram for our results.
A number of the observed sources in each bin is shown with digits. 
The scale for
the amount of modeled sources in not normalized because here 
the idea is just to
demonstrate that even by shape these two distributions are very different. The
synthetic distribution is single-peaked with a maximum close to the position
of the M7. I.e., in the theoretical model
we have a lack of cooling NSs with standard fields
$\sim10^{12}$~G (similar to normal pulsars).

\begin{figure}
\includegraphics[width=\hsize]{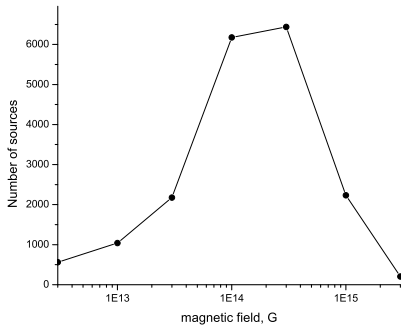}
\caption{Magnetic field distribution for modeled sources presented in Figs.1
and 2.  Only bright sources with ROSAT count rates 
$> 0.1$~cts~s$^{-1}$
contribute to the statistics. Number of sources is not normalized.
}
\label{bfield2010}
\end{figure}

This deficit is also visible in Fig.\ref{bfield2010} 
where I show the magnetic
field distribution for NSs which contribute to the population with 
relatively large
ROSAT fluxes ($\ga 0.1$~cts~s~$^{-1}$, see Fig.~\ref{lognlogs}). 
There are very few NSs with fields
$\sim 10^{12}$~G among bright sources in the synthetic model. 
On one hand, this discrepancy can be due to low statistics in the
observational data. But most probably, there is a more physical reason
behind it. For example, the synthesis can predict smaller number of 
short spin period sources if 
temperatures of standard field NSs can be underestimated in theoretical
cooling curves.

\subsection{New cooling curves}

\begin{figure}
\includegraphics[width=\hsize]{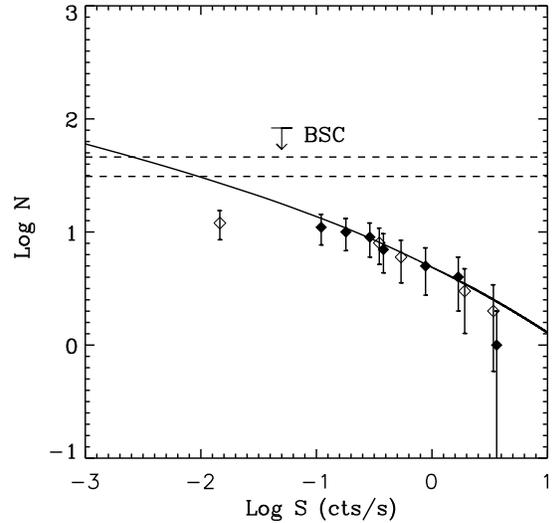}
\caption{The Log $N$ -- Log $S$ distribution for close-by NSs calculated with 
new evolutionary tracks by Vigano et al. (2013)\nocite{v2013}. 
Solid line corresponds to the results of calculations. 
The horizintal axis is for ROSAT counts per second.
Symbols represent the observed
sources. Filled symbols correspond to addition to the distribution
of one of the M7 sources, opaque --- to a radio pulsar. Poissonian errors are
shown. The BSC limit corresponds to the ROSAT Bright Source catalogue
(Voges et al. 1999\nocite{v1999}).   
}
\label{lognlogs}
\end{figure}

\begin{figure}
\includegraphics[width=\hsize]{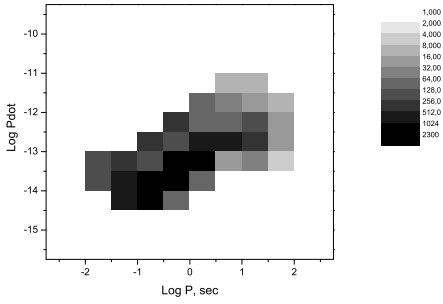}
\caption{The $p$~--~$\dot p$ distribution for the new evolutionary tracks by
Vigano et al. (2013)\nocite{v2013}. 
Numbers of sources are not normalized.
}
\label{ppdot_p10}
\end{figure}

\begin{figure}
\includegraphics[width=\hsize]{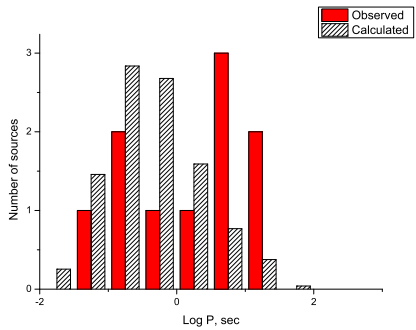}
\caption{Observed and calculated period distributions. Calculations are done
with the new evolutionary tracks by Vigano et al. (2013)\nocite{v2013}.
Total numbers of sources are equal in two distributions.
}
\label{p_p10}
\end{figure}

Vigano et al. (2013)\nocite{v2013} 
produced a different set of cooling curves, in which temperature
of objects with $B\la 3\times10^{12}$~G is higher than in the pervious variant
(used in \cite{p2010}).\footnote{I thank Daniele Vigano and Jose Pons for
the permission to use these evolutionary tracks in my calculations.} 
The results of population synthesis modeling with new tracks is shown in
Figs.\ref{lognlogs}-\ref{p_p10}. 

I used the same initial magnetic field distribution, and the same 
assumptions about the period evolution as in (\cite{p2010}). The
Log $N$~--~Log~$S$ distribution is not influenced significantly, and the
model nicely follows the data, Fig.~\ref{lognlogs}. 
But the spin period distribution for bright
sources (Fig.\ref{p_p10}) and $p$~--$\dot p$ (Fig.\ref{ppdot_p10})
plot are notably modified.

With the new cooling curves it is possible 
to increase the number of NSs with standard fields among X-ray
bright population. 
Still, spin period and  $p$~--$\dot p$ distributions are again
single-peaked, and so we do not see two separate populations of sources. 

While potentially it can be possible to fit a cooling history in such a way
that one obtains two separate populations, or at least can fit the data
taking into account low statistics of known sources, I do not think that
this is a real solution to the problem. 

\subsection{No field decay}

 One solution of the ``one second'' problem
can be related to a bimodal initial magnetic field distribution.
Such a distribution is favoured if there is an additional mechanism for
enhancing magnetic field of magnetars and other highly magnetized NSs.

I briefly illustrate this using the cooling curves by St.~Petersburg
group (\cite{casa2011}).\footnote{I thank Peter Shternin for providing me
these data.}
In this model the thermal evolution of NSs is not related to their magnetic
field evolution (and so the spin period evolution is not linked with the
thermal evolution). In the simplest case we can assume
constant fields.

\begin{figure}
\includegraphics[width=\hsize]{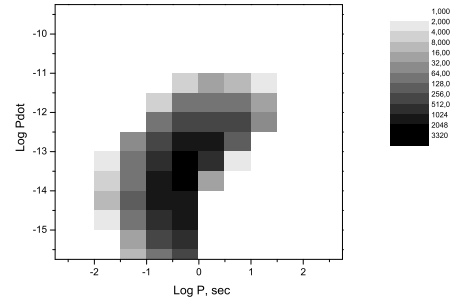}
\caption{The $p$~--~$\dot p$ distribution for constant magnetic fields and cooling
according to Shternin et al. (2011)\nocite{casa2011}. 
Magnetic field distribution is Gaussian 
with $\sigma_\mathrm{B}=0.55$ and $\langle\log B_0/\mathrm{[G]}\rangle=12.65$.
}
\label{kaspi}
\end{figure}  

At first I show $p$~--~$\dot p$ distribution for the case when 
the initial
parameters are distributed according to (\cite{fgk2006}). 
The initial magnetic field 
and the initial spin periods both  have Gaussian distributions
(for the field in log scale): 
$\langle \log B_0/\mathrm{[G]} \rangle \approx 12.65$,
$\sigma_\mathrm{B_0}\approx 0.55$ and   
$\langle p_0/\mathrm{[s]}  \rangle \approx 0.3$, 
$\sigma_\mathrm{p_0}\approx 0.15$.  

The Log $N$~--~Log $S$ distribution can be obtained in correspondence with
observations for realistic NS mass spectrum (now it is crucial, as there is
no additional heating due to field decay, and cooling is determine by
internal NS structure and processes). 
As expected, the $p$~--~$\dot p$ distribution is single-peaked.

\begin{figure}
\includegraphics[width=\hsize]{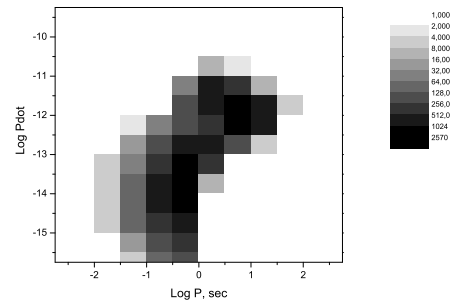}
\caption{The $p$~--~$\dot p$ distribution for constant magnetic fields and
cooling 
according to Shternin et al. (2011)\nocite{casa2011}. 
Magnetic field distribution consists of
two Gaussians. The low-field one provides 60\% of NSs and has
$\sigma_\mathrm{B}=0.5$ and $\langle \log B_0/\mathrm{[G]}\rangle=12.5$. 
The high-field
part has $\sigma_\mathrm{B}=0.15$ and $\langle \log B_0/\mathrm{[G]}\rangle=14$. 
}
\label{kaspi2}
\end{figure}  

Just for an illustration, I also produce a plot for a double-peaked field
distribution, Fig~\ref{kaspi2}. Note, that here no additional heating is added. 
In this artificial distribution the low-field part contributes 60\%
of NSs and has $\langle \log B_0/\mathrm{[G]} \rangle \approx 12.5$,
$\sigma_\mathrm{B_0}\approx 0.5$. The high-field distribution has
$\langle \log B_0/\mathrm{[G]} \rangle \approx 14$,
$\sigma_\mathrm{B_0}\approx 0.15$. 
As before only sources with ROSAT count rate $\ge 0.1$~cts~s$^{-1}$
contribute to the statistics. 

As we can see, obviously it is possible to fit the data playing with the
initial magnetic field distribution. The same can be done in the model with
decaying field. 

\section{Discussion}

In this note it is demonstrated that the bimodal period distribution
of near-by cooling NSs cannot be explained if their initial field
distribution is single-mode unless some 
unrealistic fitting assumptions are made.   
The population synthesis is not able to probe the situation deeper because
models of thermal and magneto-rotational evolution are not very certain in
several aspects. Only 
progess in understanding the NS physics might help to
solve this problem.

 The explanation of the dip around $\sim 1$~s in the period distribution by
a bimodal field distribution can be justified by proposals of a
separate mechanism of field enhancement in magnetars (\cite{td1993}), and by the
possibility of separate evolutionary channels which produce them
(\cite{pp2006}). Otherwise, we have to assume that a single smooth initial
field distribution in the range
from $\la10^{11}$~G up to $\sim10^{15}$~G is due to a
unique mechanism, which is doubtful.

``Grand unification of NSs'' is not complete until central compact objects
(CCOs) are included in the general picture. Now the most promising
approach is related to the idea of re-emerging magnetic field (\cite{ho2011,
vp2012}, Bernal, Page \& Lee 2013\nocite{bpl2013}). 
In this scenario magnetic field of a NS is submerged in
an episode of fall-back accretion. Such objects can appear as
CCOs -- low-field thermal emitters without radio pulsar activity. 
Then on time scale $\sim10^4$~yrs
this field diffuses out. Objects which passed this stage potentially can
contribute to the local population of coolers, as their fraction is
estimated to be high -- up to 30\%.  How the existence of such objects
influences the spin period distribution of close-by cooling NSs is not
known.

Finally, discussing near-by cooling NSs we are still dealing with low
statistics.  Since ROSAT time no new 
near-by candidates have been reported (see
\cite{a2011} for a recent attempt to identify more cooling NSs). 
Only more distant candidates are discussed,
and their nature is still somehow uncertain (\cite{pires2012}).
I hope very much that in near future more close-by cooling NSs
can be discovered, in the first place thanks to eROSITA on-board
Spectrum-RG. The prediction is that new discoveries will confirm the dip
around one second in the spin period distribution of near-by coolers.

\section{Conclusions}

 Here I presented the results on population synthesis modeling of period and
$p$~--~$\dot p$ distributions for young cooling NSs in the solar vicinity. 
It is shown that both calculated 
distributions are single mode for a single mode initial
magnetic field distribution. 
This does not fit observational data. The number of observed objects is not
large, however as they not only form two subpopulations in both distributions,
but also the observational properties of sources in each mode are different
(radio pulsars vs. M7), I take this dichotomy as a serious
fact.

The problem can be solved either via 
fine tunning of thermal and magneto-rotational
evolution of NSs, or via assuming bimodal initial field distribution. 
The second possibility seems to be more realistic. In this case the
M7 becomes more closely related to magnetars, i.e. their initial fields
might be enhanced similarly to magnetar fields. 

\acknowledgements
I thank Jose Pons and Daniele Vigano for the opportunity to use evolutionary
tracks calculated with their model. Also I thank Peter Shternin and Dmitry
Yakovlev for cooling tracks for NSs with constant field. Special thanks to
Andrei Igoshev for discussions and comments on the manuscript. 
This study was supported by the RSF (grant 14-12-00146). 
The author is the ``Dynasty'' foundation fellow.



\end{document}